\documentclass[twocolumn,showpacs,preprintnumbers,amsmath,amssymb]{revtex4}
\usepackage{graphicx}
\usepackage{amsmath}
\begin{document}

\title{Current reversals in a rocking ratchet: the frequency domain}

\author{A. Wickenbrock$^{1}$, D. Cubero$^{2}$,
N.A. Abdul Wahab$^{1}$, P. Phoonthong$^{1}$, and F. Renzoni$^{1}$}

\affiliation{$^{1}$Department of Physics and Astronomy, University College
London, Gower Street, London WC1E 6BT, United Kingdom}

\affiliation{$^{2}$
Departamento de F\'{\i}sica Aplicada I, EUP, Universidad de Sevilla,
Calle Virgen de \'Africa 7, 41011 Sevilla, Spain and
F\'{\i}sica Te\'orica, Universidad de Sevilla, Apartado de Correos
1065, Sevilla 41080, Spain}

\date{\today}

\begin{abstract}
Motivated by recent work [D.~Cubero {\it et al.}, Phys. Rev. E
{\bf 82}, 041116 (2010)], we examine the mechanisms which determine 
current reversals in rocking ratchets as observed by varying the frequency 
of the drive.  We found that a class of these current reversals in the 
frequency domain are precisely determined by dissipation-induced symmetry 
breaking.  Our experimental and theoretical work thus extends and generalizes
the previously identified relationship between dynamical and symmetry-breaking
mechanisms in the generation of current reversals.
\end{abstract}


\maketitle

\section{Introduction}

In out-of-equilibrium systems, directed transport can be obtained 
without the application of a net bias force. Such a counter-intuitive 
phenomenon is usually termed the ratchet effect 
\cite{comptes,magnasco,adjari,bartussek,doering,cubero06,reimann,rmp09}.
The occurrence of a directed current in such systems can be precisely related 
to the breaking of the relevant spatio-temporal symmetries 
\cite{flach00,flach01,super}.

An intriguing feature of ratchets are current reversals, where the sign of 
the generated current changes following a variation in a system parameter
\cite{bartussek,jung,mateos}.
As a specific example, consider the case of a rocking ratchet, consisting of 
Brownian particles in an asymmetric sawtooth potential. A sinusoidal rocking 
force drives the system out of equilibrium, thus allowing the generation
of directed transport. In such a system, current reversals can be observed by 
varying the noise strength at fixed amplitude of the rocking force, as well as
by varying the force amplitude while keeping constant the noise strength
\cite{bartussek}.  The key feature of current reversals is that the sign of 
the current can be reversed by varying a system parameter, although the 
considered parameter does not change the symmetry of the Hamiltonian. At first
sight this suggests that current reversals are a dynamical phenomenon, not 
traceble back to a symmetry-breaking mechanism. 

Previous theoretical work aimed to identify the mechanisms underlying 
current reversals in rocking ratchets. For an underdamped deterministic 
ratchet, Mateos \cite{mateos} argued that current reversals induced by a 
variation of the driving strength correspond to a bifurcation from a chaotic
to a periodic regime. However, the proposed mechanisms turned out not to be 
general, as in the very same system current reversals can be observed also in 
the absence of such bifurcations, and moreover not all chaos-to-order 
transitions necessarily lead to current reversals \cite{barbi,anatole}.  

A different approach toward the understanding of current reversals induced 
by a variation of the rocking force amplitude was recently introduced for a 
biharmonically driven spatially symmetric rocking ratchet \cite{cubero}. 
It was shown that a class of current reversals is precisely determined 
by symmetry breaking. In this way, a link was established between dynamical 
and symmetry-breaking mechanisms.  The still open issue now is to which 
extent the above link can be generalized, either to different systems or to 
current reversals of different type  in the same set-up. 

In the present work, we generalise the link between dynamical and 
symmetry-breaking mechanisms introduced in the aforementioned work. We 
consider the same ratchet system and examine the current reversals induced by 
a variation of the frequency of the rocking force.  We notice that a 
variation of the rocking force frequency cannot be mapped onto a variation of 
the force amplitude, i.e. the current reversals considered here are not a 
priori equivalent to the ones observed by varying the force strength.  
Our experimental and theoretical work shows that also in this case there is 
a class of current reversals that are determined by symmetry breaking,
thus generalizing the argument put forward in Ref.~\cite{cubero}. 

This work is organized as follows. In Section II we introduce the ratchet 
set-up used in this work, and recall the elements of the symmetry analysis
that will be needed to establish the link between current reversals in the
frequency domain and dissipation-induced symmetry breaking. Section III 
presents experimental results obtained with cold atoms in a driven optical 
lattice in the regime of weak damping. Section IV analyses theoretically the 
relationship between dynamical and symmetry-breaking mechanisms for the 
current reversals induced by a variation in the driving frequency.
Conclusions are drawn in Sec. \ref{sec:conclusions}.

\section{Set-up and symmetries}

\subsection{Ratchet set-up} 

In this work we consider a ratchet set-up consisting of Brownian particles
in a spatially symmetric periodic potential driven by a time-asymmetric force.
To capture the main features of 
the dynamics, in the theoretical analysis we consider the simple case of a 
linear friction, as usually considered in the ratchet literature. This will 
also allow us to explore different regimes of damping and noise.
The relevant Langevin equation is in this case:
\begin{equation}
m\ddot{x}=-\alpha\dot{x}-U'(x)+F(t)+\xi(t)
\end{equation}
where $U(x)=U_0\cos(2kx)/2$ is a periodic potential, $\alpha$ is the
friction coefficient, $\xi(t)$ is a Gaussian white noise:
$\langle \xi(t)\rangle = 0$, $\langle \xi(t)\xi(t')\rangle = 2D\delta(t-t')$,
and $F(t)$ is a biharmonic drive described by:
\begin{equation}
F(t) = F_0\left[ A_1\cos(\omega t)+ A_2\cos(2\omega t+\phi)\right]~.
\label{eq:drive}
\end{equation}

The appearance of a ratchet-effect, i.e. the generation of directed motion,
in such a set-up is a very well established fact 
\cite{fabio,chialvo,dykman,goychuk,luchinsky2,machura}. 

\subsection{Symmetry analysis}

We now recall the essential
elements of the symmetry analysis that, initially developed to explain 
the generation of a current \cite{flach00,flach01,super}, will allow us to 
introduce the basic concepts that will be needed to establish the link 
between current reversals in the frequency domain and dissipation-induced 
symmetry breaking.

In general, the symmetry analysis \cite{flach00,flach01,super} is used to 
identify the symmetry of the system which prevent directed motion. For the 
specific case of a spatially symmetric potential, of interest here, there are 
two of those symmetries: the shift symmetry, which corresponds to invariance 
under the transformation $(x,p,t)\to (-x,-p,t+T/2)$, with $T$ the period of 
the drive, and the time-reversal symmetry, which requires invariance under 
the transformation $(x,p,t)\to (x,-p,-t)$. A bi-harmonic drive of the 
form of Eq.~\ref{eq:drive} breaks the shift symmetry for any value of the 
relative phase $\phi$. For the time-reversal symmetry it is necessary to 
distinguish different cases, corresponding to different levels of dissipation.
For no dissipation (Hamiltonian case), the system is symmetric under 
time-reversal for $\phi=n\pi$ with $n$ integer, and therefore for these values
directed motion cannot be produced. It can be shown \cite{flach00,niurka} that
the dependence of the average velocity on the phase $\phi$ is, in leading 
order, $v = A\sin\phi$. Consider now the case of non zero dissipation. For
non zero dissipation, the time-reversal symmetry is broken by dissipation
also for $\phi=n\pi$ with $n$ integer. Thus, a current can be generated also
for these values of the phase $\phi$. For weak dissipation, the dependence of
the average velocity on the phase $\phi$ is, in leading order, 
$v = A\sin(\phi-\phi_0)$, where $\phi_0$  is a dissipation-induced 
symmetry-breaking phase lag which vanishes in the Hamiltonian limit 
\cite{flach01,gommers,niurka}. Finally, in the overdamped regime, the system is 
invariant under the so-called "supersymmetry" \cite{super} 
$(x,p,t)\to (x+\lambda/2,-p,-t)$, with $\lambda$ the spatial period of the 
potential, for $\phi=\pi/2 + n\pi$, with $n$ integer. For these values of 
the phase $\phi$, direction motion is not allowed.

\section{Experimental results}

The experimental set-up and procedure are substantially the same as the ones 
used in our previous work of Ref.~\cite{cubero}, and we recall here only the 
essential elements. Up to $10^8$ $^{87}$Rb atoms are trapped and 
cooled down to $\sim 50$ $\mu K$ in a magneto-optical
trap. The atoms are then loaded in a 1D lin$\perp$lin dissipative optical 
lattice, in which the atom-light interaction determines both a periodic
potential for the atoms and the dissipation mechanism which leads to a friction
force and to fluctuations in the atomic dynamics.
A driving of the form of Eq.~(\ref{eq:drive}) is applied by phase modulating 
one of the lattice beams \cite{advances}. For all the measurements presented in 
this work, the ratio between harmonics is kept fixed: $A_1=2$, $A_2=1$. The 
motion of the atoms in the driven lattice is studied by imaging the atomic 
cloud with a CCD camera. The velocity of the center of mass of the atomic 
cloud is then derived by using these images.

\begin{figure}[h]
\begin{center}
\includegraphics[height=4.in]{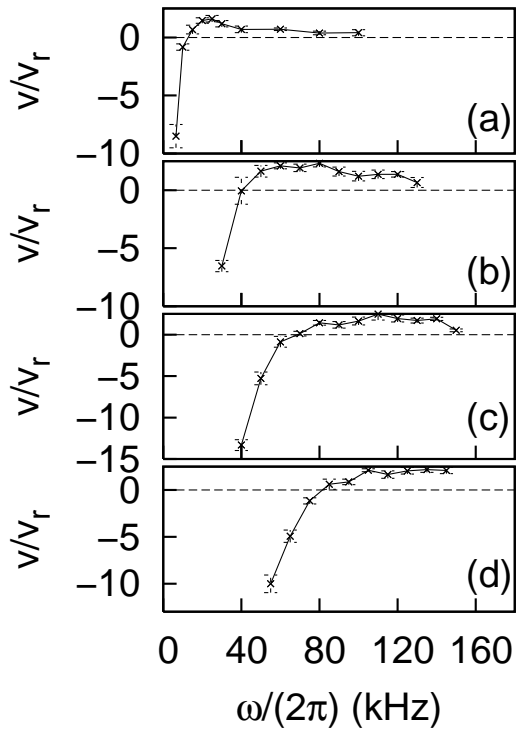}
\end{center}
\caption{
Experimental results for 1D rocking ratchet for cold atoms. The average atomic
velocity, rescaled by the recoil velocity $v_r$ ($v_r=5.88$ mm/s for $^{87}$Rb)
is reported as a function of the frequency of the drive, for a fixed value of
the relative phase between harmonics of the ac drive. The different data sets
correspond to different amplitudes of the driving force. The parameters of the
optical lattice are: detuning from resonance $\Delta=-9\Gamma$ and intensity 
per lattice beam: $I_L=(43.5\pm 0.3)$ mW/cm$^2$. The driving is a biharmonic
force of the form of Eq. \protect(\ref{eq:drive}), with the relative phase kept
fixed at $\phi=\pi/2$ for the presented sets of measurements. The amplitude of 
the force is  $F_0 = - m\lambda g_0$ where $m$ is the atomic mass, $\lambda$ 
the laser field wavelength, and the values for $g_0$ (in kHz$^2$) are, for 
the different set of data reported in the figures: (a) $g_0=3.2\cdot 10^3$, 
(b) $g_0=12.8\cdot 10^3$, (c) $g_0=19.2\cdot 10^3$, (d) $g_0=25.6\cdot 10^3$. 
The lines are a guide for the eye.
} 
\label{fig:fig_exp1}
\vspace{2em}
\end{figure}

We first consider the standard configuration for the detection of current 
reversals: for a fixed Hamiltonian, with broken symmetry, the current is 
studied as a function of a parameter, whose variation does not change the
symmetry of the Hamiltonian.  In the present case, we fix the ralative phase
between driving harmonics $\phi=\pi/2$, so to break the time reversal 
symmetry, and study the current as a function of the driving frequency
$\omega$. Results of our experiment are shown in Fig.~\ref{fig:fig_exp1}.
The familiar situation of current reversals in the frequency domain
\cite{luchinsky2,gommers05} is observed: by varying the frequency of the 
drive it is possible to revert the direction of the atomic current through 
the lattice. 


We now study the atomic current as a function of the phase $\phi$ for different
values of the driving frequency, in a range of frequencies around the value
at which the current reversal is observed. This will allow us to establish a 
relationship between the observed current reversal and dissipative effects.

\begin{figure}[h]
\begin{center}
\includegraphics[height=5.in]{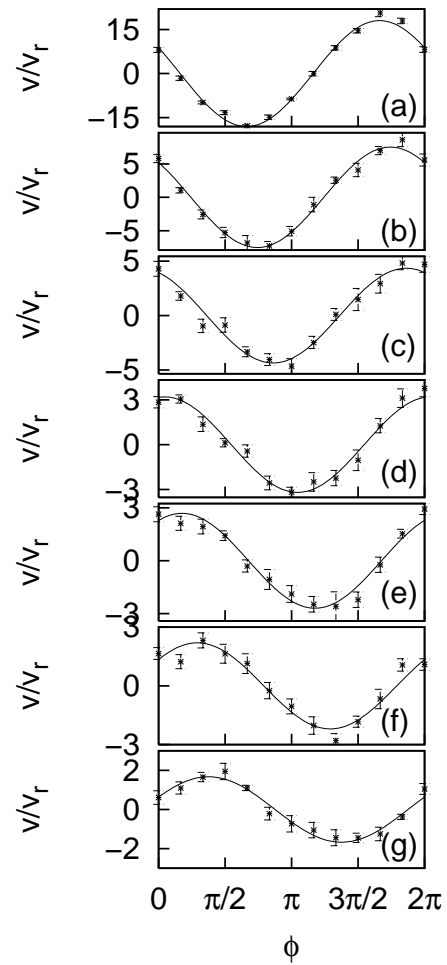}
\end{center}
\caption{
Experimental results for 1D rocking ratchet for cold atoms. The driving is 
a biharmonic force of the form of Eq.~\protect(\ref{eq:drive}). The average 
atomic velocity, rescaled by the recoil velocity $v_r$, is plotted as a 
function of the relative phase $\phi$ between harmonics of the ac drive,
for different values of the driving force frequency. The lines are the best
fit of the data with the function $v/v_r = A \sin(\phi-\phi_0)$. The lattice
parameters are the same as in Fig.~\protect\ref{fig:fig_exp1}. The frequency
of the drive for the different data sets is:
(a) $\omega/(2\pi)= 40$ kHz, (b) $\omega/(2\pi)= 50$ kHz,
(c) $\omega/(2\pi)= 60$ kHz, (d) $\omega/(2\pi)= 70$ kHz,
(e) $\omega/(2\pi)= 80$ kHz, (f) $\omega/(2\pi)= 100$ kHz,
(g) $\omega/(2\pi)= 120$ kHz. The driving force amplitude, for all data
sets, is $F_0=-m\lambda g_0$ with $g_0=19.2\cdot 10^3$ kHz$^2$.
}
\label{fig:fig_exp2}
\vspace{2em}
\end{figure}

The results of our measurements are shown in Fig.~\ref{fig:fig_exp2}, for
a driving force amplitude corresponding to that of Fig.
\ref{fig:fig_exp1}(c). For all considered driving frequencies, the dependence
of the current on the phase $\phi$ is well described by
$v/v_r=A\sin(\phi-\phi_0)$, with $\phi_0$ a dissipation-induced
symmetry-breaking phase lag.  This is in agreement with previous observations
\cite{gommers,cubero}. The important, and so far unexplored, fact here
is that $\phi_0$ varies significantly when the driving frequency is scanned
across the value corresponding to the current reversal in the frequency domain.

\begin{figure}
\hspace*{-3.5cm}\includegraphics[height=3.in]{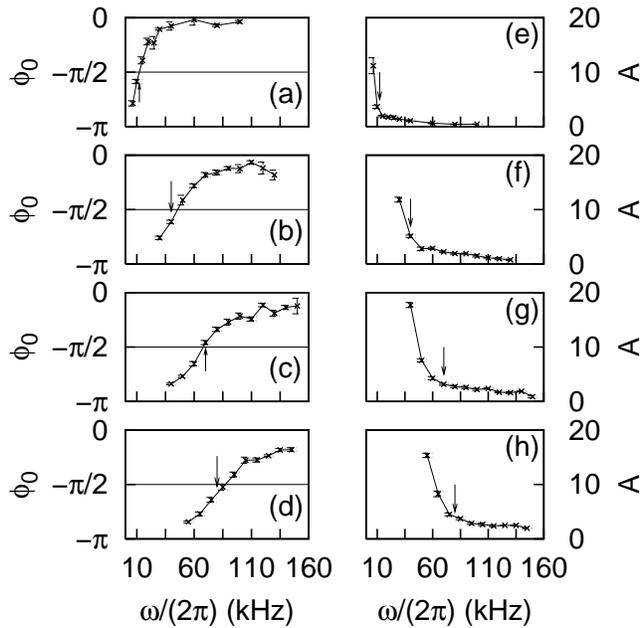}
\caption{
Left (a-d): dissipation-induced phase lag $\phi_0$, as obtained by fitting 
data as those in Fig.~\ref{fig:fig_exp2}, with the function 
$v/v_r = A \sin(\phi-\phi_0)$, as a function of the frequency of the drive, 
for different values of the driving strength. 
Right (e-h): amplitude A, obtained from the same fits.
Different rows of plots of the figure correspond to different driving 
strength, with the amplitudes of the drive the same as in Fig.
\protect\ref{fig:fig_exp1}, e.g. the data in (a) and (e) correspond
to a driving amplitude $g_0=3.2\cdot 10^3$ kHz$^2$ and so on.
The lattice parameters are the same as in Fig.~\protect\ref{fig:fig_exp1}.
The lines are a guide ot the eye. The arrows indicate the frequency 
at which the current reversal shown in Fig.~\protect\ref{fig:fig_exp1} occurs.
}
\label{fig:fig_exp3}
\vspace{2em}
\end{figure}

The dependence of the phase-lag $\phi_0$ and the amplitude $A$, as obtained by
fitting data as those of Fig.~\ref{fig:fig_exp2} with the function
$v/v_r = A \sin(\phi-\phi_0)$, on the driving frequency is reported in
Fig.~\ref{fig:fig_exp3}. The four different data sets correspond to the
four different driving strength amplitudes for which the data in Fig.
\ref{fig:fig_exp1} were obtained. The phase-lag $\phi_0$ shows a very
large variation around the frequency at which the current reversal is
observed in Fig.~\ref{fig:fig_exp1}, with the phase lag varying from $-\pi$ 
to $0$ around the current
reversal, and taking the value $-\pi/2$ at the reversal frequency \cite{note}. 
Since in the experiments reported in Fig.~\ref{fig:fig_exp1} the relative phase is fixed to $\phi=\pi/2$, this value of the phase-lag, $\phi_0=-\pi/2$, guarantees that the current is suppressed at the reversal frequency, where the directed current inverts its sign. 
On the other hand, the amplitude of the sin-like curve, see right panels in Fig.~\ref{fig:fig_exp3}, stays finite around the reversal frequency. Thus, we can conclude that the
current reversal in the frequency domain is determined by the large variation,
around the reversal frequency, of the dissipation-induced symmetry-breaking
phase lag $\phi_0$. This generalizes the link between symmetry-breaking
and current reversals established in Ref.~\cite{cubero} in the case of
current reversals in the amplitude domain.

\section{Numerical analysis}

The presented experimental results show that the current reversals in the 
frequency domain are determined by a sharp variation of the dissipative
phase lag $\phi_0$ when the driving frequency is varied around the value
corresponding to the current reversal.

The conditions of our experiment correspond to the weakly damped regime,
hence the appearance of a nonzero dissipation-induced phase lag $\phi_0$.
The validity of the established link between current reversals and 
dissipation-induced symmetry breaking can be tested by considering the 
Hamiltonian and the overdamped limit. In both cases, the phase lag 
$\phi_0$ is fixed by the symmetries of the system valid in the respective
limits: $\phi_0=0$ for the Hamiltonian limit and $\phi_0=-\pi/2$ \cite{note2}
for the 
overdamped case, as discussed in Sec. II.B. Thus $\phi_0$ cannot vary with 
the driving frequency, and the current reversals observed in the case of 
moderate dissipation should disappear.

Our experimental set-up is not suitable to explore the two extreme case of no-dissipation,
and very large dissipation. Thus, in order to show that these current 
reversals vanish when the phase lag $\phi_0$ is fixed by the symmetries in 
the Hamiltonian and overdamped regimes, we resort to numerical simulations 
of Eq.~(1).

\begin{figure}[h]
\vspace{1.cm}
\begin{center}
\includegraphics[height=3in]{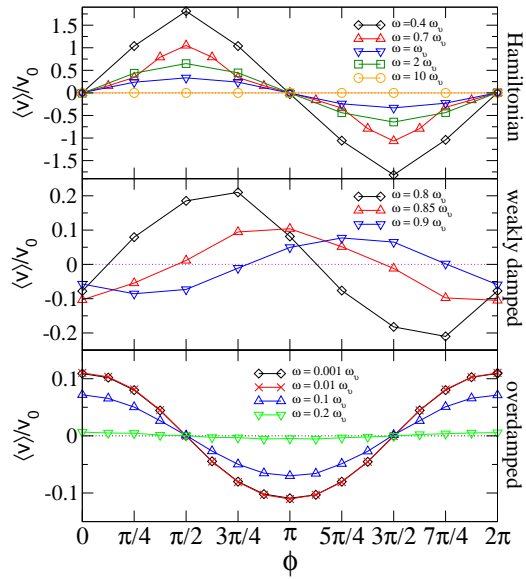}
\end{center}
\caption{(Color online) Average atomic velocity as a function of the relative
phase $\phi$ between harmonics of the ac drive for several values of the 
driving frequency $\omega$, with $\omega_v=k(2U_0/m)^{1/2}$ and $A_1=A_2=1$.
Top panel: simulation data in the Hamiltonian regime ($\alpha=D=0$) for 
$F_0=0.2 \,U_0k$.
Middle panel: simulation data in the weakly damped regime. The driving 
amplitude, the friction and the noise strength values are fixed to 
$F_0 = 0.2 \,U_0k$, $\alpha=0.15 \alpha_0$, and $D=1.944 D_0$, respectively,
where $\alpha_0=m k v_0$, $D_0=\alpha_0^2 v_0/k$, and $v_0=(U_0/m)^{1/2}/10$.
The values about $\phi=\pi/2$ (or $\phi=3\pi/2$) show a current reversal as 
the driving frequency is increased.  Bottom panel: simulation data in the 
overdamped regime for $F_0= U_0k$, $\alpha=100 \alpha_0$, and 
$D=1.944\cdot 10^3 D_0$. Lines are a guide to the eye.}
\label{fig:fig_theory}
\end{figure}

Let us examine first the weakly damped regime with this model. The middle
panel of Fig. 4 shows the average particles' velocity as a function of the
driving phase for different values of the driving frequency. A current reversal
can be clearly observed at about $\phi=\pi/2$ (or $\phi=3\pi/2$) as the driving frequency is increased, being a
consequence of a variation of the phase lag $\phi_0$ with the driving
frequency. This is the regime explored in our experiment.

On the other hand, the phase lag $\phi_0$ is fixed to $0$ in the Hamiltonian
regime due to the time-reversal symmetry. Accordingly, the top panel of Fig.~4
shows that
no current reversals are observed when the driving frequency is varied. In the
limit of strong dissipation, the appearance of the supersymmetry fixes the
phase lag to $\phi_0=-\pi/2$, as shown in the bottom panel of
Fig.~\ref{fig:fig_theory}. This fact prevents the occurrence of the current
reversals observed in the presence of moderate dissipation.

The numerical simulations thus confirm the validity of the established link 
between current reversals in the frequency domain and dissipation-induced
symmetry breaking.

\section{Conclusions} \label{sec:conclusions}

In conclusion, we examined the mechanisms which determine
current reversals in rocking ratchets as observed by varying the frequency
of the drive. We considered the specific case of a bi-harmonically driven 
spatially symmetric ratchet. We found that a class of these current reversals 
in the frequency domain are precisely determined by dissipation-induced 
symmetry breaking. Correspondingly, these reversals are observed only for 
moderate dissipation, and disappear in the Hamiltonian and overdamped limit.

Our experimental and theoretical work thus extends and generalizes
the previously identified relationship \cite{cubero} between dynamical 
and symmetry-breaking mechanisms in the generation of current reversals.

In the future, it would be interesting to re-examine the issue of the 
relationship between microscopic dynamics and current reversals in the  
system studied by Mateos \cite{mateos}. Whether the current reversals 
discussed there can be associated to some distinctive features of 
the microscopic dynamics is still an open question. 

\acknowledgements
This research was supported by the Leverhulme Trust. One of us (DC)
also thanks the Ministerio de Ciencia e Innovaci\'on of Spain for financial
support (grant FIS2008-02873).


\begin{thebibliography}{99}
\bibitem{note} Not all the plots in the figure show exacly $\phi_0=\pi/2$. This
is due to a measurement error, as $\phi_0$ is determined by fitting the 
entire sin-like curve obtained by varying $\phi$ while the reversal frequency 
is determined by measuring only the current for $\phi=\pi/2$.
\bibitem{note2} Note that the fact that the phase lag is $\phi_0=-\pi/2$ and not $\phi_0=\pi/2$ (assuming a positive $A$) cannot be determined a priori by the symmetry analysis, and here is reported based on the a posteriori observation of the numerical results.
\end{thebibliography}
\end{document}